\documentclass{PoS}
\usepackage{picinpar}

\title{$b \to d$ Penguins: CP Violation, General Lower Bounds 
 on the Branching Ratios and Standard Model Tests}

\ShortTitle{$b\to  d$ Penguins: CP Violation, General Lower Bounds 
 and Standard Model Tests}

\author{Robert Fleischer
\hspace*{8.5cm}\parbox{3cm}{\rm\normalsize 
hep-ph/0511325\\\mbox{CERN-PH-TH/2005-237}\\TUM-609/05\\[-1.5cm]}\\
        CERN, Switzerland\\
        E-mail: \email{Robert.Fleischer@cern.ch}}

\author{\speaker{Stefan Recksiegel}\\ 
        TU Muenchen, Germany\\
        E-mail: \email{Stefan.Recksiegel@ph.tum.de}}

\abstract{With the wealth of new data from the $B$-factories,
$b\to d$ penguin decays become available for study, in addition
to their $b\to s$ counterparts that have proven an
indespensable tool for the exploration of new-physics effects
in flavour physics. A prominent example of the $b\to d$ penguin 
transitions is $\bar B^0_d \to K^0 \bar K^0$. We show that this decay 
can be charaterized in the Standard Model by a
surface in the observable space of the direct and mixing-induced CP
asymmetries and the branching ratio. The form of this surface, which
is theoretically clean, implies a lower bound for the branching ratio that has 
recently been confirmed
experimentally. If future measurements of the CP asymmetries yield a
point away from the SM surface, this would be an interesting signal of new physics.
We point out that the hadronic parameters in $\bar B^0_d \to K^0 \bar K^0$
that parameterize the position on the SM surface are related to hadronic
parameters in the $B \to \pi K$ system. The fact that the branching ratio of
$\bar B^0_d \to K^0 \bar K^0$ is very close to its lower bound yields
interesting implications for $B \to \pi K$ even without knowledge of the
CP asymmetries of $\bar B^0_d \to K^0 \bar K^0$.
The mechanism that produces the lower bound for $\bar B^0_d \to K^0 \bar K^0$
is actually much more general; we derive lower bounds for various other
$b \to d$ penguin-induced processes, including $B \to \rho \gamma$ and $B^\pm
\to K^{(\ast)\pm} K^{(\ast)}$. 
Some of these theoretical lower bounds are very close to the current experimental
upper bounds. }

\FullConference{International Europhysics Conference on High Energy Physics\\
		 July 21st - 27th 2005\\
		 Lisboa, Portugal}

\begin{document}

\section{Introduction}
Flavour-changing neutral-current (FCNC) processes, possible in the Standard Model
(SM) only through loop diagrams, are an extremely important probe for new
physics (NP). The good agreement between experiment and theory in
processes induced by $b\to s$ FCNCs has already put important  
constraints on physics beyond the SM. Due to the excellent work of the
$B$-factories, we are now entering the era where $b\to d$ penguin-induced
processes -- typically suppressed by a factor of 20 with respect to
the corresponding $b\to s$ penguin transitions -- can be used to test the SM
more rigorously than it was possible before.

The flavour structure of the SM, more specifically the order of magnitude of the 
individual elements of the Cabibbo--Kobayashi--Maskawa (CKM) matrix, allows us
to derive certain relationships between different observables in 
$b\to d$-induced decays, and between $b\to s$- and $b\to d$-related
observables. These relationships allow us to test the SM in those cases
where the corresponding observables have already been measured and
to make predictions where observations are still missing.

\section{$B_d^0\to K^0\bar K^0$: CP Violation and the Branching Ratio}

In the SM, we can write the amplitude for the decay $B_d^0\to K^0\bar K^0$ as
\begin{equation}\label{ampl-BdKK}
A(B_d^0\to K^0\bar K^0)=\lambda^{(d)}_u {\cal P}_u^{K\!K} + 
\lambda^{(d)}_c {\cal P}_c^{K\!K} +  \lambda^{(d)}_t {\cal P}_t^{K\!K},
\end{equation}
where the $\lambda^{(d)}_q \equiv V_{qd}V_{qb}^\ast$ are CKM factors,
and the ${\cal P}_q^{K\!K}$ denote the strong amplitudes of penguin topologies
with internal $q$-quark exchanges, which receive tiny contributions from
colour-suppressed electroweak (EW) penguins and are fully dominated by 
QCD penguin processes. Eliminating $\lambda^{(d)}_t$ with the
help of the unitarity relation $\lambda^{(d)}_t=-\lambda^{(d)}_u-\lambda^{(d)}_c$
of the CKM matrix, we can write the amplitude as
\begin{equation}\label{ampl-BdKK-lamt}
A(B^0_d\to K^0\bar K^0)=\lambda^3A{\cal P}_{tc}^{K\!K}
\left[1-\rho_{K\!K} e^{i\theta_{K\!K}}e^{i\gamma}\right],
\end{equation}
where  ${\cal P}_{tc}^{K\!K}\equiv {\cal P}_t^{K\!K}-{\cal P}_c^{K\!K}$,
and $\rho_{K\!K} e^{i\theta_{K\!K}}$ is a function of the 
${\cal P}_q^{K\!K}$ that we treat as an unknown hadronic
parameter.

\begin{window}[0,l,{\vspace{1mm}
\includegraphics[width=7.5cm]{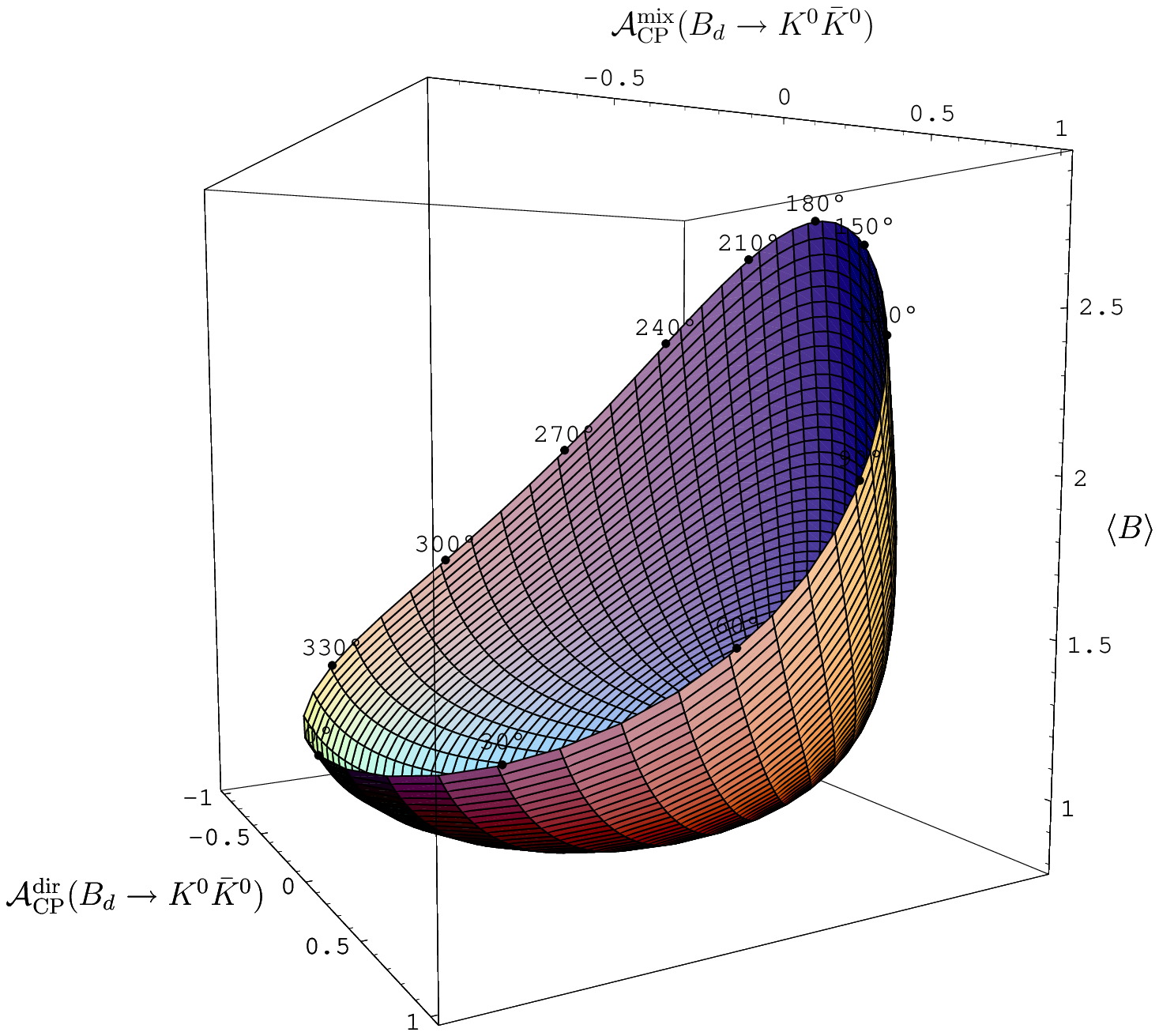} \vspace{0mm}},{
{\bf Figure 1:} The surface in observable space.}]
The direct and mixing-induced CP asymmetries
${\cal A}_{\rm CP}^{\rm dir}(B_d\to K^0\bar K^0)$ and
${\cal A}_{\rm CP}^{\rm mix}(B_d\to K^0\bar K^0)$
are functions of {\it only} $\rho_{K\!K}$, $\theta_{K\!K}$,
the angle $\gamma$ of the unitarity triangle, and (in the latter case) the
$B^0_d$--$\bar B^0_d$ mixing phase $\phi_d$; the same
is true for the normalized branching ratio $\langle B \rangle$,
where phase-space and CKM factors as well as $|{\cal P}_{tc}^{K\!K}|^2$
have been factored off. 

For fixed values of $\gamma$ and $\phi_d$,
$\rho_{K\!K}$ and $\theta_{K\!K}$ then span a surface in the
${\cal A}_{\rm CP}^{\rm dir}$--${\cal A}_{\rm CP}^{\rm 
mix}$--$\langle B \rangle$  observable space, 
shown in  Fig.~1 for 
$\phi_d = 47^\circ$ and $\gamma=65^\circ$.  (The fringe 
is defined by $\rho_{K\!K}=1$, the numbers give the
value for $\theta_{K\!K}$.)
\end{window}

In the SM, any
measurement of the three observables has to lie on this surface,
which is theoretically clean. 
Sufficiently accurate measurements of the branching ratio will
give strong constraints on possible values for the asymmetries.

The form of the surface implies a theoretical {\it lower}
bound for $\langle B \rangle$ that can be converted into a
lower bound for $\mbox{BR}(B_d\to K^0\bar K^0)$ using input
from $b\to s$ penguin decays (see \cite{FR1} for details). 
With the help of this lower bound, the recent measurement of
$B_d\to K^0\bar K^0$ \cite{BK0K0exp} was correctly
predicted in \cite{FR1}. Using the latest experimental input
and the central values of the factorizable $SU(3)$-breaking
parameters, we update the bound in (3) of \cite{FR1} to
${\rm BR}(B^0_d \to \bar K^0 K^0)> 1.43\,^{+0.17}_{-0.25}$, 
nicely consistent with the old result and the recent
measurements (see Table~\ref{BRtable}).

We observe that the measured ${\rm BR}(B^0_d \to \bar K^0 K^0)$
is right at the lower theoretical bound (bottom of the surface
in Fig.~1). This implies a value of $\rho_{K\!K}$ significantly
different from 0, with a small phase $\theta_{K\!K}$; $\rho_{K\!K}$
can be related to a hadronic $B\to \pi K$ parameter through 
$\rho_{\rm c}=\epsilon \rho_{K\!K}$, where 
$\epsilon\equiv\lambda^2/(1-\lambda^2)=0.053$. This quantity
is usually neglected. However, a value of $\rho_{\rm c}\sim 0.05$, as 
suggested by ${\rm BR}(B^0_d \to \bar K^0 K^0)$, would be rather 
welcome in the analysis of the $B\to \pi K$ system~\cite{UPDATE}.

\section{General Lower Bounds on the Branching Ratios of
$b\to d$ Penguin Processes}

The mechanism that provided the lower bound on
${\rm BR}(B^0_d \to \bar K^0 K^0)$ is actually more general.
We will now first use it to derive lower bounds on $b\to d \gamma$ processes,
and then discuss the general $b\to d$ penguin case.
The amplitude for the decay $\bar B \to \rho\gamma$ can be written as
\begin{equation}\label{Ampl-Brhogam}
A(\bar B \to \rho\gamma)=c_\rho \lambda^3 A {\cal P}_{tc}^{\rho\gamma}
\left[1-\rho_{\rho\gamma}e^{i\theta_{\rho\gamma}}e^{-i\gamma}\right],
\end{equation}
where $c_\rho=1/\sqrt{2}$ and 1 for $\rho=\rho^0$ and $\rho^\pm$,
respectively, and $A=|V_{cb}|/\lambda^2$. Moreover,
${\cal P}_{tc}^{\rho\gamma}\equiv{\cal P}_t^{\rho\gamma}-{\cal P}_c^{\rho\gamma}$, 
where ${\cal P}_t^{\rho\gamma}$ and ${\cal P}_c^{\rho\gamma}$ are matrix
elements of operators from the standard weak effective Hamiltonian
(see \cite{FR2} for details). $\rho_{\rho\gamma}e^{i\theta_{\rho\gamma}}$ is 
again a hadronic parameter that we will treat as essentially unknown. 
Let us now use the information offered by the $b\to s$ counterpart of our 
$b\to d$ transition, which is well measured and takes an amplitude of the 
following form:
\begin{equation}\label{Ampl-BKastgam}
A(\bar B \!\to\! K^\ast \!\gamma)\!=-\!
\frac{\lambda^3 \! A {\cal P}_{tc}^{K^\ast\!\gamma}}{\sqrt{\epsilon}} \!
\left[1\!+\!\epsilon\rho_{K\!^\ast\!\gamma}e^{i\theta_{K\!^\ast\!\gamma}}
e^{-i\!\gamma}\right]\!,
\end{equation}
where $\epsilon$ was introduced above.  The ratio of the corresponding BRs
is then given by
\begin{equation}\label{rare-ratio}
\frac{\mbox{BR}(\bar B \to \rho 
\gamma)}{\mbox{BR}(\bar B \to K^\ast \gamma)}=\epsilon
\left[\frac{\Phi_{\rho\gamma}}{\Phi_{K\!^\ast\gamma}}\right]
\left|\frac{{\cal P}_{tc}^{\rho\gamma}}{{\cal P}_{tc}^{K\!^\ast\gamma}}
\right|^2 H^{\rho\gamma}_{K\!^\ast\gamma},
\end{equation}
where $\Phi_{\rho\gamma}$ and $\Phi_{K\!^\ast\gamma}$ denote phase-space 
factors, and 
\begin{equation}
H^{\rho\gamma}_{K\!^\ast\gamma}\equiv
\frac{1-2\rho_{\rho\gamma}\cos\theta_{\rho\gamma}\cos\gamma+
\rho_{\rho\gamma}^2}{1+2\epsilon\rho_{K\!^\ast\gamma}
\cos\theta_{K\!^\ast\gamma}
\cos\gamma+\epsilon^2\rho_{K\!^\ast\gamma}^2}.
\end{equation}
Although $\rho_{K\!^\ast\gamma}e^{i\theta_{K\!^\ast\gamma}}$ is here
strongly suppressed by $\epsilon$, we can straightforwardly include the
corresponding corrections by using the flavour-symmetry relation
$\rho_{\rho\gamma}e^{i\theta_{\rho\gamma}}=
\rho_{K\!^\ast\gamma}e^{i\theta_{K\!^\ast\gamma}}\equiv \rho e^{i\theta}$. 
Treating then $(\rho,\theta)$ as completely free parameters, we can derive
the following lower bound:
\begin{equation}\label{H-bound}
H^{\rho\gamma}_{K\!^\ast\gamma}\geq \left[1-2\epsilon
\cos^2\gamma+{\cal O}(\epsilon^2)\right]\sin^2\gamma,
\end{equation}
which can be converted into a lower bound for $\bar B \to \rho \gamma$
through (\ref{rare-ratio}) and the measured value of the $\bar B \to K^\ast \gamma$
branching ratio.
\vspace*{3mm}

\begin{window}[0,l,{\vspace*{0mm}
\includegraphics[width=7.5cm]{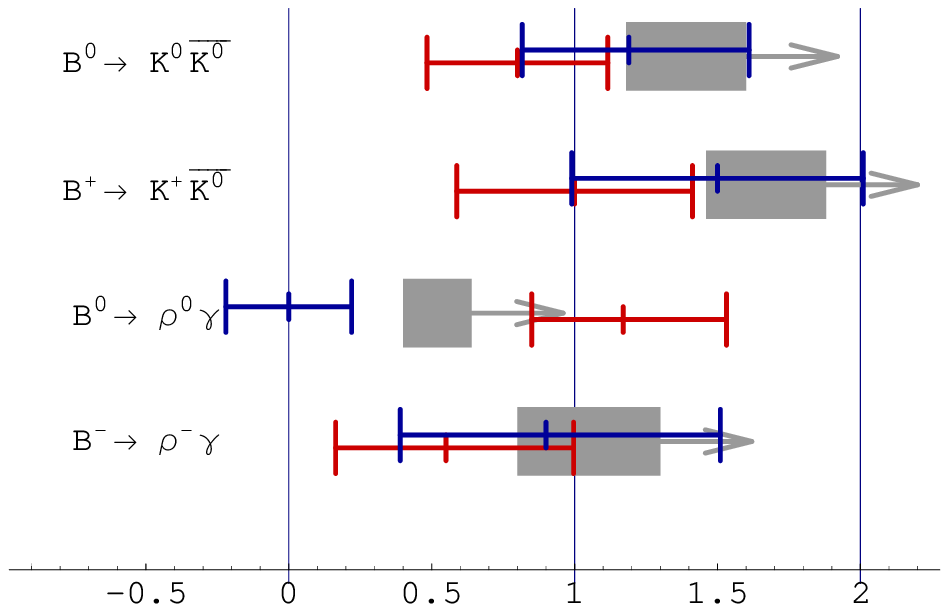} \vspace{0mm}},{
{\bf Figure 2:} Experimental results (upper: BaBar, lower: Belle)
and theoretical limits}]
Taking into account phase-space effects and factorizable
$SU(3)$-breaking corrections, we obtain the lower bounds
given in Table~1. For comparision, we also show the bounds
that result from neglecting the $SU(3)$-breaking corrections
(``na\"ive bound''). The bounds are consistent with the
experimental results for $B^0_d \to \rho^0 \gamma$ and
 $ B^+ \to \rho^+ \gamma$, although of course the well-known 
 isospin-breaking puzzle of the Belle result remains.

In a similar way we can also derive theoretical lower limits
for other $b\to d$ penguin decays. We list bounds for
$B^\pm\to K^{(\ast)\pm} K^{(\ast)}$ -- together with the respective
$b\to s$ decay that was used for the bound -- in Table 1
(experimental data are taken from \cite{HFAG}); some
more channels, including also $B^\pm \to \pi/\rho^\pm \ell^+ \ell^-$
modes, are discussed in \cite{FR2}. For the currently most
interesting decays, the theoretical predictions and measurements
by BaBar and Belle are also plotted in Fig.~2. It will be interesting
to confront our bounds with future data.
\end{window}

\begin{table}
\begin{tabular}{|c|c||c|c|c|c|c|} \hline
 $b\to s$ process & Exp.\ rate & $b\to d$ process & Na\"ive bound & Bound &
 Belle & BaBar \\
 \hline
$B^+ \to K^0 \pi^+$ & $24.1\pm1.3$ & $ B^0_d \to \bar K^0 K^0$
 & $ 0.88\,^{+0.11}_{-0.15}$ & $ 1.43\,^{+0.17}_{-0.25}$ & $ 0.8\pm 0.32$ & $ 1.19\,^{+0.42}_{-0.37}$\\
\hline
$B^+ \to K^0 \pi^+$ & $24.1\pm1.3$ & $ B^+ \to \bar K^0 K^+$ & $
1.03\,^{+0.11}_{-0.14}$ &$ 1.69\,^{+0.19}_{-0.23}$&  $ 1.0\pm 0.41$ & $ 1.50\pm 0.51$\\
\hline
$B^+ \to K^{\ast 0} \pi^+$ & $9.7\pm 1.2$ & $B^+ \to \bar K^{\ast 0} K^+$ &
$0.46\,^{+0.06}_{-0.07}$ &$0.76\,^{+0.10}_{-0.12}$ & \multicolumn{2}{|c|}{$<5.3$ (CLEO)}\\
$B^+ \to K^{\ast 0} \rho^+$ & $10.6\pm1.9$ & $B^+ \to \bar K^{\ast 0} K^{\ast
  +}$ & $0.46\,^{+0.09}_{-0.10}$ &$0.73\,^{+0.15}_{-0.16}$&
  \multicolumn{2}{|c|}{$<71$ \,(CLEO)}\\
\hline
$B^0_d \to K^{\ast 0} \gamma$ & $40.1\pm2.0$ & $ B^0_d \to \rho^0 \gamma$ &
 $  0.86\,^{+0.10}_{-0.12}$& $ 0.51\,^{+0.13}_{-0.11}$ & $
  1.17\,^{+0.36}_{-0.32}$ & $0.0\pm0.22$ \\
\hline
$B^+ \to K^{\ast +} \gamma$ & $40.3\pm2.6$ & $ B^+ \to \rho^+ \gamma$ & 
 $ 1.73\,^{+0.22}_{-0.26}$& $ 1.03\,^{+0.27}_{-0.23}$ & $
  0.55\,^{+0.45}_{-0.39}$ & $ 0.9\,^{+0.61}_{-0.51}$ \\
\hline
\end{tabular}
\caption{ Post-LP2005: { Belle} and { BaBar} { data}, {theory lower bounds.}
\label{BRtable}}
\end{table}

\end{document}